\documentstyle[12pt]{article}
\begin{document}

\title{\bf DELTA DECAY IN NUCLEAR MEDIUM \\}

\author{\bf B.K.Jain and Bijoy Kundu \\
Nuclear Physics Division \\
Bhabha Atomic Research Centre, Bombay-400 085, India.\\}
\date{}

\maketitle

\begin{abstract}
Proton-nucleus collisions, where the
beam proton gets
excited to the delta resonance and then decays to p$\pi ^+$,
either inside or outside the nuclear medium, are studied. Cross-sections
for various kinematics for the (p,p$' \pi ^+$) reaction
between 500 MeV and 1 GeV beam energy are calculated to see the
effects of the nuclear medium on the propagation and
decay of the resonance.
The cross-sections studied
include proton energy spectra in coincidence with the pion,
four momentum transfer distributions, and
the invariant p$\pi^+$ mass distributions. We find that the effect of
the nuclear medium on these cross-sections mainly reduces their
magnitudes. Comparing these cross-sections
with those considering the decay of the delta outside
the nucleus only, we further find
that at 500 MeV
the two sets of cross-sections have large differences, while by 1 GeV the
differences between them become much smaller.

\end{abstract}
PACS number(s): 25.40.Ve, 13.75.-n, 25.55.-e
\newpage

\section{\bf INTRODUCTION}
In the description of nuclear reactions at intermediate energy,
in addition to nucleons and pions, deltas play an important role.
Because of this, there has been great interest over the years in
the study of the delta-nucleus interaction [1]. Experimentally, since
the delta is a spin-isospin excitation of the nucleon, it is readily
excited in charge-exchange reactions such as (p,n) and ($^3$He,t).
Measurements of inclusive spectra on ejectiles in these reactions
show broad bumps with large cross-sections around 300 MeV excitation
in all targets ranging from $^{12}$C to $^{208}$Pb [2]. These
excitations
correspond to the excitation of a nucleon in the target to a delta
isobar. They are, therefore, capable of exploring delta-nucleon hole
($\Delta N^{-1}$)
excitations in nuclei and thereby the collective aspects of these
modes.

Another class of intermediate energy reactions, such as (p,$\Delta ^{++}$),
are those
where the $\Delta $ appears as one of the final reaction products.
In these reactions the measurements can be done directly on the delta
or its decay products. They are, therefore, well suited
to investigate delta dynamics in the continuum and
the transition interaction $pp \rightarrow n\Delta ^{++}$. The
presence of a delta in these reactions can be inferred either by
measuring the recoiling nucleus,
as in the pioneering experiment on the $^6$Li(p,
$\Delta ^{++}$)$^6$He reaction at Saturne [3], or the ejectile nucleus
on proton targets, as in the experiments on p($^3$He,t)$\Delta ^{++}$
and p($^{12}$C,$^{12}$B)$\Delta ^{++}$ reactions carried out at
Saturne and Dubna [2,4]. These kind of experiments, however, can only be
performed on few nuclei as the ejectile nucleus is required to be
stable or sufficiently long lived. Theoretical analyses of these
reactions, which
consider the delta as a stable elementary particle, show that this
reaction proceeds in one step and the measured cross-sections can
be adequately described in the framework of the distorted-wave
Born approximation (DWBA) [5]. It has also been found that, due to the
very large momentum transfer ($\sim $ beam momentum) involved
in the excitation of the bound  nucleon in the target to the delta
in the
continuum in the final state, the `target excitation' in the
(p,$\Delta ^{++}$) contributes little [6].

Yet another way to study the (p,$\Delta ^{++}$) reaction is to
detect the $\Delta $ directly
by measuring its decay products p and $\pi ^+$. This procedure has the
great advantage that the experiment can be performed on any
target nucleus. Furthermore, by choosing a different
energy of the outgoing delta we can vary the energy transfer to the
nucleus, and thereby explore its spin-isospin response at different
excitations. Experiments on (p,p$'\pi ^+$)
are now being done at ~~~TRIUMF with the availability of
dual magnetic spectrometers [7]. For analyzing these
reactions, it is necessary to develop a formalism which, unlike
earlier work, incorporates the
unstable nature of the delta and includes the effect of the nucleus on
its propagation
and on the decay products p$'$ and $\pi ^+$. In this work we
present such a formalism. Then using it, we calculate
various cross-sections which can be measured on the (p,p$' \pi ^+$)
reaction. We study the effects on these reactions of the nuclear distortion
of the delta. We compare these
cross-sections with those calculated with the delta decaying outside
the nucleus only. This comparison determines the region of applicability
of the latter model, where the calculations can be done with much
ease and where the delta-nucleus interaction can be explored without
the complicating effects due to the p$\pi ^+$ interaction with the nucleus.

The reaction mechanism which we follow for the (p,p$' \pi ^+$) reaction
is shown in fig. 1. Accordingly, the incoming proton interacts
with a nucleon at some point $\bf r$ and is converted into a delta.
The target nucleon, to provide one unit of spin and
isospin to the beam proton,
undergoes a spin and isospin flip. It also gets
accelerated by a momentum corresponding to the excitation
energy ($\approx $300 MeV) of the delta.
The delta then propagates
from the point $\bf r$ to $\bf r'$, where it
decays into a proton and pion. The decay point $\bf r'$ may lie
inside or outside the nucleus. During this propagation, the delta
interacts with the nuclear medium and suffers distortion.
The incoming proton and the decay products p$'$ and
$\pi ^{+}$ get distorted by the nuclear medium
before reaching the
point $\bf r$  and in propagating out of the point
$\bf r'$, respectively. In our formalism, we include all these
distortions through appropriate optical potentials. The formalism
itself is written
following Gottfried and Julius [8], who,
among several other workers [9], have studied the effect of
the nuclear medium on the decay of a $\rho $ -meson.

For the elementary process pp $\rightarrow$ n$\Delta ^{++}$ in the
above mechanism we have used the
one-pion-exchange potential. We have not included the contribution
from the $\rho $ exchange for this excitation. The reason for this
omission is that
attempts to include $\rho $ exchange in the
pp $\rightarrow$ n$\Delta ^{++}$ reaction have yielded very
unsatisfactory results. A detailed study
by Jain and Santra [10] and earlier work by Dmitriev et al. [11]
both showed that the experimental data on
the spin averaged cross-sections for this reaction
are better reproduced by the one pion-
exchange interaction alone. The inclusion of rho-exchange worsens this
agreement. This means that either the strength of the
$\rho N\Delta$ coupling, f$_{\rho N\Delta}$, is considerably
weaker than what is usually assumed, or some additional amplitude tends
to cancel the contribution from the rho. A recent theoretical study on the
microscopic analysis of the $\rho N\Delta$
vertex due to Liu and Haider [12], in fact, suggests that the value of
f$_{\rho N\Delta}$ is much smaller.

In Section 2 we present the formalism in detail. Sections 3 and 4
contain results, discussion and the conclusions.
Since the contribution of the delta decay
inside the nucleus depends upon the beam energy and the size of
the nucleus we have done calculations at 500 MeV and 1 GeV
for $^{12}$C and~~ $^{208}$Pb.

In general, our findings are that
(i) around 1 GeV and for lighter target nuclei the calculated
cross-sections do not differ greatly from those which consider
the decay of
the delta only outside the nucleus, and (ii) the delta-nucleus
interaction mainly reduces the magnitude of the cross-sections.

 \section{\bf FORMALISM}

The differential cross-section for the A(p,p$'\pi^+$)B reaction is given by
\begin{equation}
d\sigma=[PS]<|T_{fi}|^2>,
\end{equation}
where [PS] is the factor associated with phase space and the beam current,
\begin{equation}
[PS]=\frac {1}{j(2 \pi)^5}\frac {m_\Delta m^2}{E_\pi E_p E_{p'}}
\delta ^4(P_i-P_f)d{\bf k}_{p'}d{\bf k}_\pi d{\bf K}_B.
\end{equation}
Here $j$ is the beam current and m is the mass of the proton. $P_x$
denotes the four momentum.

In the centre of mass system, $j$ is given by
\begin{equation}
j=\frac {p_c \sqrt s}{E_p E_A},
\end{equation}
where $\sqrt s$ is the available energy in the centre of mass system
and $p_c$ is the c.m. momentum in the initial state.

$T_{fi}$ in eq.(1) is the transition amplitude. The angular brackets
around its square
denote the appropriate sum and average over the final and
initial spins, respectively.

\subsection{\bf Transition Amplitude, $T_{fi}$}
 For the diagram in fig. 1, the transition amplitude can be written as
\begin{equation}
T_{fi}=\int d{\bf r}d{\bf r'}{\chi _{p'}({\bf r'})^-}^*
{\chi _\pi^-}^*({\bf r'})\Gamma_{\Delta N\pi}
G_\Delta ({\bf r',r})\psi_\Delta ({\bf r}),
\end{equation}
where $\Gamma_{\Delta N\pi}$ is the decay operator for the delta
decaying into
p$\pi ^+$. In momentum space and in a non-relativistic static
approximation, it is
given by
\begin{equation}
\Gamma_{\Delta N\pi}=\frac{f_\pi^*}{m_\pi}{\bf S^\dagger .\kappa_\pi
 T^\dagger .\phi_\pi},
\end{equation}
 where $\bf S$ and $\bf T$ are the spin and iso-spin transition operators
 respectively for $\frac {1}{2} \rightarrow \frac{3}{2}$.  $\kappa $
 is the pion momentum in the delta rest frame.
Because the final pion is on-shell ( if we
neglect the effect of distortions on it), the above form for $\Gamma_
{\Delta N\pi}$
does not contain the usual
form factor F$^*$ .

G$_\Delta ({\bf r',r})$ is the delta propagator. It satisfies the inhomogeneous
wave equation
\begin{equation}
[\nabla ^2+E_p^2-m_\Delta ^2+i\Gamma _\Delta m_\Delta -\Pi_\Delta({\bf r})]
G_\Delta ({\bf r',r})=\delta ({\bf r'-r}),
\end{equation}
where m$_\Delta$ (=1232 MeV) and $\Gamma _\Delta$ are the resonance parameters
associated with the free delta. The width of the free delta, $\Gamma_\Delta$,
depends upon the invariant mass according to
\begin{equation}
\Gamma (\mu)=\Gamma_0 \Bigl[\frac{ k(\mu^2,m_\pi^2)}
{k(m_\Delta^2,m_\pi ^2)}\Bigr]^3
\frac{k^2(m_\Delta ^2,m_\pi ^2)+\gamma ^2}{k^2(\mu ^2,m_\pi ^2)+\gamma ^2},
\end{equation}
with $\Gamma _0$=120 MeV and $\gamma$=200 MeV. $\mu$ is the invariant mass
given by
\begin{equation}
\mu ^2=(E_{p'}+E_\pi)^2-({\bf k}_{p'}+{\bf k}_\pi)^2
\end{equation}
In equation (7) $k$, for an on-shell pion, is given by
\begin{equation}
k(\mu^2,m_\pi^2)=[(\mu^2+m^2-m_\pi^2)^2/4\mu^2-m^2]^{1/2}
\end{equation}
This relation reflects the restrictions on the available phase space for the
decay of a delta of mass $\mu$ into an on-shell pion of mass m$_\pi$ (=140 MeV).

$\Pi_\Delta$ in the Green's function, eq.(6), describes the collisions
of the delta with the medium. In the mean field approximation, it is related to
the delta optical potential, V$_\Delta$, through
\begin{equation}
\Pi_\Delta=2E_\Delta V_\Delta .
\end{equation}
 One of the
important channels
which contribute significantly to this potential is
$\Delta$N$\rightarrow$NN.

$ \psi_\Delta ({\bf r})$ in eq. (4) is the amplitude for the production
of the delta at a point $\bf r$. It
is given by
\begin{equation}
\psi_\Delta ({\bf r})=\int d{\bf \xi} \psi _\beta ^*({\bf \xi})
\Gamma _{\pi NN}
\psi _\alpha ({\bf \xi})G_\pi ({\bf r,x})\Gamma _{\pi N\Delta}
\chi _p^+({\bf r}).
\end{equation}
 Here $\Gamma _{\pi N\Delta}$ is the operator for the excitation of
 the beam proton to the delta,
 and  $\Gamma _{\pi NN}$ is the interaction
 operator at the
 $\pi NN$ vertex in the nucleus. Like $\Gamma _{\Delta N\pi}$, their
 forms are,
 \begin{equation}
 \Gamma_{\pi N\Delta}=i\frac{f_\pi^*F^*(t)}{m_\pi}\bf{S.qT.\phi _\pi},
 \end{equation}
 and
 \begin{equation}
 \Gamma_{\pi NN}=i\frac{f_\pi F}{m_\pi}\bf{\sigma .q \tau .\phi _\pi},
 \end{equation}
 with f$_\pi ^*$ and f$_\pi $
 the coupling constants for the $\pi N\Delta$ and $\pi NN$
 vertices. Their values are 2.156 and 1.008, respectively. F$^*$
 and F are the form factors associated with these vertices. They
 incorporate the off-
 shell extrapolation of the pion-nucleon coupling vertex.
 t is the four momentum transfer
 (squared). However, for the ``projectile excitation'' diagram,
  if we ignore the
 nuclear recoil, it is the same as the three momentum, ${\bf q}$, squared.

$\chi_p$ in eq. (4) is the distorted wave for the beam proton. In
this paper we will consider beam energies above 500 MeV. We use the
eikonal approximation for this and the other continuum
particles (viz. $\Delta, p'$ and $\pi^+$). This implies that the main effect
of the nuclear medium on these waves is absorptive, and the dominant
momentum components in them
are their asymptotic momenta. Because of this, and also because
the interaction $V_{NN\rightarrow N\Delta}$ in the
region of t of the present studies
is known to depend weakly on the momentum transfer q [5],
the evaluation of $\psi_\Delta({\bf r})$ simplifies. In the case
of closed shell nuclei we find
\begin{equation}
\psi_\Delta({\bf r})\approx
\frac{<\Gamma_{\pi NN}({\bf q})>}{(m_\pi ^2-t)}
\Gamma _{\pi N\Delta} \rho _{\beta \alpha}({\bf r})\chi _p^+({\bf r}).
\end{equation}
Here the angular brackets around $\Gamma_{\pi NN}$ represent its
expectation value over the nuclear spin-isospin wave functions.
$\rho _{\beta \alpha}({\bf r})$ is the spatial part of the nuclear
transition density.
The momentum
transfer, $\bf q$, is given in terms of the asymptotic momenta of the
continuum particles as
\begin{equation}
\bf q=k_p-k_\Delta (=k_p{'}+k_\pi)
\end{equation}

Substituting $\psi_\Delta$ from eq. (14) in eq. (4), we get
\begin{equation}
T_{fi}=\frac {<\Gamma_{pn \pi ^+}>}{q^2+m_\pi^2}<\Gamma_{ pp'\pi^+}>
F_{fi}(k_{p'},k_\pi ;k_p),
\end{equation}
where
\begin{equation}
F_{fi}=\int d{\bf r}d{\bf r'}\chi_{p'}^{-*}({\bf r'}) \chi_\pi^{-*}({\bf r'})
G_\Delta ({\bf r',r})\rho _{\beta \alpha}({\bf r})\chi_p^+({\bf r}),
\end{equation}
and
\begin{equation}
<\Gamma_{ pp'\pi^+}>=[\frac {f_\pi ^*}{m_\pi}]^2 F^*(t)
<\sigma_p'|{\bf S^\dagger .\kappa_\pi T^\dagger .
\phi_\pi S.q T.\phi_\pi}|\sigma_p>
\end{equation}

$<|T_{fi}|^2>$ is given by
\begin{equation}
<|T_{fi}|^2>=\frac {|<\Gamma_{ pp'\pi^+}>|^2|<\Gamma_{pn\pi^+}>|^2}{(
m_\pi^2-t)^2}|F_{fi}({\bf k_{p'},k_\pi ;k_p})|^2,
\end{equation}
where $|<\Gamma_{ pp'\pi^+}>|^2$ is given by
\begin{eqnarray}\nonumber
&|<\Gamma_{ pp'\pi^+}>|^2&=\frac {1}{2} \Sigma_{\sigma_p \sigma_{p'}}
|\Gamma_{ pp'\pi^+}|^2\\
& &=\frac{1}{9}[\frac {f^*}{m_\pi}]^4F^{*2}(t)[4|{\bf \kappa_\pi .q}|^2+
|{\bf q}\times {\bf \kappa}_\pi|^2].
\end{eqnarray}
In the last evaluation we have used the identity
\begin{equation}
{\bf S^\dagger .qS.\kappa}_\pi=\frac {2}{3}{\bf \kappa_\pi.q}-
\frac {i}{3}{\bf \sigma
.(q}\times {\bf \kappa}_\pi)
\end{equation}

\subsection{\bf Evaluation of F$_{fi}{\bf (k_{p'},k_\pi;k_p)}$}

To evaluate
\begin{equation}
F_{fi}=\int d{\bf r}d{\bf r'}\chi_{p'}^{-*}({\bf r'})\chi_\pi ^{-*}({\bf r'})
G_\Delta ({\bf r',r})\rho_{\beta \alpha}({\bf r})\chi_p^+({\bf r}),
\end{equation}
we define, for convenience, a function
\begin{equation}
G_\Delta ({\bf r;k}_\Delta,\mu)=\int d{\bf r'}\chi_{p'}^{-*}({\bf r'})
\chi_\pi ^{-*}({\bf r'})G_\Delta ({\bf r',r}).
\end{equation}
This function physically gives the probability amplitude for
finding a proton and a pion in the detector with the total momentum,

\begin{equation}
{\bf k}_\Delta ={\bf k_p'+k_\pi},
\end{equation}
and the invariant mass $\mu$
 if the delta
is produced at a point $\bf r$ in the nucleus. The wave equation for this
new function can be obtained from eq. (6) by multiplying it on both sides
by $\chi_{p'}^{-*}\chi_\pi ^{-*}$ and integrating over $\bf r'$. This gives
\begin{equation}
[\nabla ^2+E_p^2-m_\Delta ^2+i\Gamma_\Delta m_\Delta -\Pi_\Delta({\bf r})]
G_\Delta({\bf r;k}_\Delta,\mu)=
\chi_{p'}^{-*}({\bf r})\chi_\pi ^{-*}({\bf r}).
\end{equation}
Eq. (22) for F$_{fi}$ then becomes
\begin{equation}
F_{fi}=\int d{\bf r}G_\Delta ({\bf r;k}_\Delta,\mu)\rho _{\beta \alpha}
({\bf r})\chi _p^+({\bf r})
\end{equation}
To proceed further, as mentioned earlier, we treat all the continuum waves
in the eikonal approximation. We write
\begin{equation}
\chi_p({\bf r})=e^{i{\bf k}_p.{\bf r}}D_{{\bf k}_p}({\bf r})
\end{equation}
and
\begin{equation}
G_\Delta ({\bf r;k}_\Delta,\mu)=e^{-i{\bf k}_\Delta .{\bf r}}
\Phi ({\bf r;k}_\Delta
,\mu),
\end{equation}
where the distortion functions D and $\Phi$ are slowly varying functions
of z, the coordinate taken along the beam momentum, ${\bf k}_p$. We
have evaluated these distortion functions along ${\bf k}_p$.
\begin{equation}
D_{{\bf k}_p}({\bf r})=exp[\frac {-i}{\hbar v_p}\int _{-\infty}^z dz'
V_p({\bf b},z')],
\end{equation}
where $V_p$ is the optical potential for the beam proton. The function $\Phi$,
whose dependence on ${\bf k}_\Delta$ and $\mu$ will henceforth be suppressed
for brevity,
is obtained from eq. (25). With the eikonal approximation
this simplifies to
\begin{equation}
[G_0^{-1}(\mu)-\Pi_\Delta -2ik_\Delta \frac {\partial}{\partial z}]\Phi (b,z)
=D_{p'}(b,z)D_\pi (b,z),
\end{equation}
where
\begin{equation}
G_0(\mu)=[\mu^2-m_\Delta^2+i\Gamma_\Delta m_\Delta]^{-1}
\end{equation}
 is the
propagator for the free delta. The solution of eq. (30) gives
\begin{equation}
\Phi({\bf b},z)=\frac {1}{2ik_\Delta}\int _z^\infty \Phi _\Delta ({\bf b};
z,z')f({\bf b},z')dz',
\end{equation}
where,
\begin{equation}
\Phi_\Delta({\bf b};z,z')=exp[\frac {i}{2k_\Delta}(\mu^2-m_\Delta^2+i
\Gamma_\Delta m_\Delta)(z'-z)]exp[\frac {-i}{v_\Delta}\int _z^{z'}V_\Delta
({\bf b},z'')dz''],
\end{equation}
and,
\begin{equation}
 f({\bf b},z)=exp[-i\int _{z'}^\infty (\frac {V_{p'}({\bf b},z'')}{v_{p'}}+
 \frac {V_\pi ({\bf b},z'')}{v_\pi})dz''].
 \end {equation}
Here, as we see, $\Phi _\Delta$ describes the propagation of
the delta from
z to its decay point $z'$.
f(${\bf b},z'$) describes the same for the decay products,
p$'$ and $\pi ^+$,
of the  delta from $z'$ to the detectors. Eq. (30)
for $ \Phi({\bf b},z)$ thus gives the probability amplitude for
detecting p$ ' \pi ^+$ in the detector when the delta is
distorted by the medium from its production
point z to its decay point $z'$ and
the decay products p$'$ and $\pi ^+$ are distorted from $z'$
to the detectors.

The delta dynamics implicit in $\Phi ({\bf b},z)$ become
more transparent if we consider
a nucleus with distorting potentials having a sharp surface. For a
radius R of this surface, we can split the ``distorted''
delta propagator $\Phi$ into two parts, i.e.
\begin{equation}
\Phi ({\bf b},z)=\Phi _{in}({\bf b},z)+\Phi _{out}({\bf b},z),
\end{equation}
where
\begin{equation}
\Phi _{in}({\bf b},z)=\frac {1}{2ik_\Delta }\int _z^{\sqrt (R^2-b^2)}
dz'\Phi _\Delta ({\bf b};z,z')f({\bf b},z')
\end{equation}
and
\begin{eqnarray}\nonumber
&\Phi _{out}({\bf b},z)&=\frac {1}{2ik_\Delta}\int _{\sqrt (R^2-b^2)}^\infty
dz'\Phi _\Delta ({\bf b}:z,z')f({\bf b},z')\\
& &=\frac {1}{2ik_\Delta}\int _{\sqrt (R^2-b^2)} ^\infty dz'\Phi _\Delta
({\bf b};z,z')
\end{eqnarray}
These two functions, as we see, describe respectively the
contributions to the cross-section from the decay of the delta inside
and outside the nuclear medium.
The relative contribution of these two regions, as seen
from eq.(33) for $ \Phi _\Delta({\bf b};z,z')$,
is determined by the intrinsic decay length,
\begin{equation}
\lambda _{in}=\frac {k_\Delta}{m_\Delta \Gamma _\Delta}=\tau v_\Delta
\end{equation}
of the delta. Due to this factor, the delta, in travelling a distance,
$L(\bf b,z)[=\sqrt(R^2-b^2)-z]$
from its production point z to the nuclear surface, gets attenuated  by
a factor $exp(-L(b,z)/\tau v_\Delta$). This attenuation, as is
obvious, decreases with an increase in the delta (hence the beam)
momentum and its life time. Consequently,
the ratio, $\Phi _{in}/\Phi _{out}\rightarrow 0$, as the beam momentum,
k$_p \rightarrow \infty $ and/or $\tau \rightarrow \infty$.

In case there is no distortion by the nuclear medium,
$\Phi$ can be shown to be independent
of ($\bf b$,z) and reduce to the free delta propagator, i.e.
\begin{equation}
\Phi({\bf b},z)=[\mu ^2-m_\Delta ^2+i\Gamma _\Delta m_\Delta]^{-1}\equiv G_0
\end{equation}
The effect of the nuclear field on the delta and its decay products, as is obvious
from here, results in
the modification of this mass distribution, and through it,
the modification of other experimental
observables associated with p$'$ and $\pi ^+$.

For no distortions, eq. (17) for the transition amplitude integral, F$_{fi}$,
simplifies to
\begin{equation}
F_{fi}= \rho _{\beta \alpha}({\bf q}) G_0,
\end{equation}
where
\begin{equation}
\rho _{\beta \alpha}({\bf q})=\int d{\bf r}e^{i{\bf q.r}} \rho _{\beta \alpha}
({\bf r})
\end{equation}
is the nuclear transition density in  momentum space. This expression
demonstrates that, apart from G$_0$ (or its nuclear modified version),
F$_{fi}$ and the cross-sections are determined
by the nuclear transition density.

\section{\bf RESULTS AND DISCUSSION}
The results
presented here are motivated by two aims:
(i) to see how these results compare with those in a model
which considers
the decay of the delta only outside the nucleus, and (ii)
to see the extent to which the delta distortions affect the
cross-sections. Since the outcome of both these investigations depends on
the speed of the delta in the nucleus and the distance it travels through
the nucleus, we present calculations for 500 MeV and
1 GeV beam energies and for $^{12}$C and $^{208}$Pb target nuclei.
The specific cross-sections which we calculate are the four momentum
transfer distributions, invariant mass spectra of p$\pi ^+$ and
the proton energy spectra in coincident p and $\pi ^+$
measurements.
\subsection{INPUT QUANTITIES}
The calculations require the following quantities as
input:
(i) the pion-nucleon coupling constants and form factors
at the vertices; (ii) the resonance
parameters of the delta;
(iii) the nuclear transition density, $\rho _{\beta \alpha}$, and (iv)
the optical potentials for protons, pion and delta.

The pion coupling and the delta resonance parameters have already
been given
in Section 2. These parameters reproduce the scattering data on
the $p\pi ^+ \rightarrow p\pi ^+$ [13] reaction. For the form factors
F and F$^*$, we have used the mono-pole form,
\begin{equation}
F(t)=\frac {\Lambda^2 -m_\pi ^2}{\Lambda ^2-t},
\end{equation}
with $\Lambda $=1.2 GeV/c for both the vertices.

For nuclear transition densities we have used forms which reproduce
electron scattering data. Since in the present paper our
emphasis is not on the investigation of nuclear structure
effects on the (p,p$'\pi ^+$)
cross-sections, this choice of transition densities should be adequate.
For $^{12}$C we take,
\begin{equation}
\rho_{\beta \alpha}(r)=\rho_0 [1+a(r/b)^2]exp(-r^2/b^2),
\end{equation}
and for $^{208}$Pb we choose,
\begin{equation}
\rho_{\beta \alpha}(r)=\frac {\rho_0}{1+exp((r-c)/d)}.
\end{equation}
where $\rho_0$ is fixed by the appropriate normalization of the
density.
Other parameters are taken from the
electron scattering analyses compilations
by Jackson and Barrett  and De Jager et al. [13]. They are:\\
\begin{center}
$a$=1.247, $b$=1.649 fm;\hspace{.5in} $c$=6.54 fm, $d$=0.5 fm.\\
 \end{center}

For optical potentials, since we expect their effect to be mostly
absorptive  at intermediate energies [15],
we use only their imaginary parts. For pions we fix them using
the method of Ericson
and H$\ddot{u}$fner [16], which is valid around the delta resonance.
In this method, the
strength of the optical potential is obtained from the refractive
index, $n$, of the pion in the nuclear medium, where
\begin{equation}
n(E)=\frac {K (E)}{k(E)}
\end{equation}
and
\begin{equation}
W(E)=-\frac {k^2}{E}n_I(E).
\end{equation}
Here $n_I$ is the imaginary part of the refractive index and K is the
pion wave number in the nuclear medium. The latter is obtained by solving the
dispersion relation
\begin{equation}
K ^2=k^2+4\pi \rho f_{\pi N}(K ,E),
\end{equation}
where $f_{\pi N}$ is the $\pi -$N scattering amplitude in the nuclear medium
in the forward direction. Considering that the pion scattering is dominated
by p-wave scattering, the expression for $\pi ^+$ scattering from a
nucleus with N neutrons and Z protons is
\begin{equation}
f_{\pi ^+N}=\frac {1}{A}(Nf_{\pi ^+n}+Zf_{\pi ^+p})\approx [\frac
{N+3Z}{3A}]f_{33}
\end{equation}
A Breit-Wigner resonant form for the amplitude f$_{33}$ gives
\begin{equation}
n_I(E)=\frac {X\Gamma /4}{(E-E_R+\frac {3}{4}X)^2+\frac {1}{4}\Gamma ^2},
\end{equation}
with
\begin{equation}
X=4\pi n_0 c,
\end{equation}
and
\begin{equation}
c= -[\frac {N+3Z}{3A}]\frac {58(MeV)a^3}{1+(ka)^2}.
\end{equation}
Here $n _0$ is the nuclear density, and $a$=1.24 fm.

The dispersion relation given in eq.(47) is equivalent to an optical
 potential approach,
if the latter is defined through the folding of the $\pi$-N t-matrix
with the
nuclear density. Tandy et al. [17] have shown that such optical
potentials contain
nucleon knock-out as the primary reactive content. Therefore, the main
contribution to n$_I$(E) in eq. (49) is the nucleon knock-out
($\pi ^+,\pi ^+$N) channel.
However, in addition to this, the pion flux can also be lost through
 real absorption of the pion in the nuclear medium. The
 dominant channel which contributes to such an absorption is $\Delta $N
 $\rightarrow $ NN. We have approximated this contribution to the pion
 optical potential by
 \begin{equation}
 W_{abs}=g\Gamma _s/2,
 \end{equation}
 where $\Gamma_s$ is the spreading width. We take $\Gamma_s$=70 MeV
 from the delta-hole model of pion absorption
  [18].
 The factor $g$ is added to account for the fact that the (p,p$'\pi ^+$)
 reaction is a peripheral reaction. This factor represents the
 fraction of the
 central nuclear density in the region where this reaction
 takes place. We have
 chosen $g$=0.7.
 The
 resulting pion optical potentials
 are listed in Table I.

For protons, the distorting potentials beyond 300 MeV are obtained
using the high energy ansatz,
\begin{equation}
W=\frac {1}{2}v\sigma _T^{pN}n_0,
\end{equation}
where $\sigma _T^{pN}$ is the total proton-nucleon cross-section at the proton
speed, $v$, in their c.m. The values for the total cross-section are
taken from the experimental data on proton-nucleon scattering [19].
The radial shapes, $\rho (r)/\rho (0)$, of the potentials are taken
to be the same as those given in eqs.(43,44), with n$_0$
=0.17 fm $^{-3}$. Below 300 MeV the potentials are taken from those
available in the literature from the analyses of elastic
scattering data of protons on $^{12}$C at various energies [20].
The radial
shape for these potentials is the two parameter Woods-Saxon form.
The optical potentials for $^{208}$Pb are taken of the same form except that the
radius parameter for it is enhanced in proportion to A$^{1/3}$.

On delta optical potentials, not much information exists. For T$_\Delta
\leq $100 MeV we make recourse to the delta-hole model of Hirata et al.
[18]
and take
W$_\Delta$=-45 MeV. For higher energies, as for protons, we make the high
energy ansatz as given in eq. (54). Here, however, for obtaining
$\sigma _T^{\Delta N}$
first we write it as a sum of the elastic and the reactive parts,
$\sigma _T^{\Delta N}=\sigma _{el}^{\Delta N}+\sigma _r^{\Delta N}$.
Then assuming that the spin averaged elastic dynamics of the proton and delta
are not very different, we assume $\sigma _{el}^{\Delta N}\approx
\sigma _{el}^{NN}$. For the reactive part, since up to about 1.5 GeV the
main reactive channel in $\Delta$N scattering is $\Delta $N$\rightarrow$ NN,
using the reciprocity theorem we write
\begin{eqnarray}\nonumber
&\sigma _r^{\Delta N}&\approx \sigma ^{\Delta N\rightarrow NN}\\
& &=\frac {1}{2} \frac {k_{NN}^2}{2k_{\Delta N}^2}\sigma (pp\rightarrow
n\Delta ^{++}),
\end{eqnarray}
where k$_{NN}$ is the proton c.m. momentum  corresponding
to the same energy
as is available in the $\Delta$N c.m. An extra factor 1/2 is introduced to
account for the identity of two particles in the final state.
Using the experimentally known cross-sections for the pp$\rightarrow$
n$\Delta ^{++}$ reaction [21], the resulting
delta optical potentials at representative energies are also
listed in Table I.

\subsection{CALCULATED CROSS-SECTIONS}
In figs. 2-3 we show the four momentum transfer distributions for
$^{12}$C at 500 MeV and 1 GeV beam energies.
 Theoretically, the cross-sections
are calculated by integrating eq.(59) over the delta mass between
1120 - 1300 MeV. Experimentally, these cross-sections can be obtained
either by measuring the four momentum of the recoil nucleus or
the four momenta of the decay products, p$'$ and $\pi ^+$, of the
delta. Results in figs. 2-3 have three curves. The curve
`a'
corresponds to a situation where we assume no nuclear distortion of
continuum particles. These are the plane wave (PW)
results. Curve `b' includes the
distortion of the beam proton and the propagating delta. This
curve, therefore, corresponds to a situation when the delta
is assumed to decay always outside the nucleus. Curve `c' includes,
in addition, the nuclear distortion of
p$'$ and $\pi^+$. From these curves it is evident that
d$\sigma$/dt, at both energies,
can get modified in a major way by nuclear distortions.
However, at 1 GeV the main modifications
in the PW curves arise due to proton and delta distortions;
at 500 MeV distortion of the
outgoing proton and pion is equally important. We also notice that
at higher energy the main effect of the distortions is the
reduction in the magnitude of the cross-sections. At lower energy
distortions also fill in the minima in the d$\sigma$/dt distributions.
Quantitatively, the peak cross-section at 1 GeV gets reduced by
a factor of about 6 due to the beam proton and delta distortions
and by an additional factor of 3/2 due to p$'$ and $\pi ^+$
distortions. At 500 MeV the corresponding factors are about 2
and 3 respectively.

In figs. 4-5 we display the calculated invariant mass ($\mu$)
distributions for the decay products, p$'$and $\pi ^+$, of the
delta. The results again are given for 500 MeV and 1 GeV
beam energies and the delta going in the forward direction.
For an isolated delta these distributions should peak
around 1230 MeV. In our results, all the distributions
are shifted
towards a lower mass. At 1 GeV the peak
appears around 1200 MeV, while at 500 MeV it appears around 1150 MeV.
Since this shift appears in the PW results also, it is caused primarily
by the nuclear transition density, $\rho _{\beta \alpha}$.
The distortion of the continuum particles controls the magnitude
of the cross-sections. The quantitative reduction in the
peak cross-sections at both energies are similar to those seen
earlier in the four momentum transfer distributions.
This again means that
at 1 GeV the additional effect of the p$'$ and $\pi ^+$ distortion
is not much, while at 500 MeV it is of the same magnitude as that
due to the beam and delta distortions.

Since the inclusive measurements normally tend to have large
background, it is sometimes preferable to make exclusive
measurements, though  the cross-sections
in these measurements are smaller. In
figs.6-7 we show the calculated proton energy spectra
for the coincident measurements of the decay products, p$'$ and $\pi ^+$.
As an example, the results are given for the two
particles being produced at 10 $^0$ on either side of the beam direction.
In these results we again find that at 1 GeV the dominant effect arises
due to the
distortion of the beam and delta. The distortion of p$'$ and $\pi ^+$
has a small effect. At 500 MeV, on the other hand, the distortions
of all the continuum particles have equal effect.

To see the dependence of the above results on nuclear size we
have calculated distributions for a $^{208}$Pb target.
Since for this nucleus we find that the calculated cross-sections
are very small
at 500 MeV beam energy, only 1 GeV
results are shown. In figs. 8-9 we display the
calculated invariant mass distribution and the out-going proton
energy spectrum for the coincident p$'$ and $\pi ^+$ measurements.
Unlike the $^{12}$C results, here we find that, even at 1 GeV, the effect of
p$'$ and $\pi ^+$
distortion is as important as that of the beam proton and delta.
Furthermore, the distortions here not only reduce the magnitude
of the cross-sections, they also change their shapes. The plane wave
invariant mass distribution, which previously had three peaks,
has only one peak when all the distortions are included.

Finally, In fig. 10-11  we present some results to isolate the
effect  on the
measured cross-sections due to the delta-nucleus interaction alone, and
also to see the effect of the real part of its potential on the
cross-sections. As an illustration we show
d$\sigma$/d$\mu$ for $^{12}$C at 500 MeV and 1 GeV.
The results are shown for three situations. Curves `a' include no
delta distortion, `b' include only the imaginary
part, W$_\Delta$, of V$_\Delta$,
and `c' includes both real and imaginary parts of V$_\Delta$.
The beam,
p$'$ and $\pi ^{+}$ distortions are included in all the curves.
However, unlike earlier curves, these distortions now include the
effect of their real potentials (U) also.
From the comparison of various curves in these figures, we find
that the effect of
the delta-nuclear collisions does get reflected in the final results
to a significant extent. At 1 GeV, the term W$_\Delta$ suppresses the peak
cross-section by
about 30 $\%$, while at 500 MeV this suppression factor is 50 $\%$.
The effect of U$_\Delta$ on the cross-sections, however,
is not much. At 1 GeV its effect on the magnitude of the cross-
sections is within 10 $\%$, while
at 500 MeV it is insignificant. Inclusion of U$_\Delta$ also does not lead to
any perceptible shift in the peak position of the invariant mass
distributions.

The real parts of the optical potentials for protons and deltas
in the above figures are fixed using the same procedure as given earlier for
their imaginary parts.
For pions, like the imaginary part, they are obtained through the
real part of its refractive index, i.e.
\begin{equation}
\frac {U_\pi}{E}=1-n_r^2,
\end{equation}
where
\begin{equation}
n_r(E)=1-\frac {\frac {1}{2}X(E-E_R+\frac {3}{4}X)}
{(E-E_R+\frac {3}{4}X)^2+\frac {1}{4}\Gamma ^2}.
\end{equation}

\section{Conclusions}
In conclusion, for the (p,p$'\pi ^+$) reaction,
proceeding through a delta excitation in the intermediate state,

(i) around 1 GeV beam energy and in lighter target nuclei
(like $^{12}$C), the cross-sections obtained in a model, which
considers the
delta decay only outside the nuclear medium, are not significantly
different from those obtained in a model which also includes the delta
decay inside
the nuclear medium. At lower energies ($\sim$ 500 MeV) and/or for
heavier nuclei (like $^{208}$Pb) the situation is different. The
calculated cross-sections in two models can differ significantly in
magnitude as well as shape.

(ii) The distortion of the delta by the nuclear
medium yields a suppression of the magnitude of the measured
cross-sections. The  peak position of the invariant mass and
other distributions remain uneffected by it.

(iii) The peak positions in the invariant mass and other distributions
are determined by the range of momentum transfer involved  and the
nuclear transition density. Compared to an isolated delta,
the peak positions in the $\mu$-distributions  get shifted towards
lower mass. At 1 GeV this shift is small ($\mu_{peak}$-1232
$\approx$30 MeV), while at 500 MeV it is large ($\mu_{peak}$-1232
$\approx$80 MeV).
\section{Acknowledgements}
The authors wish to thank the referee and Prof. J. T. Londergan
for many useful comments
and their assistance in improving the presentation of the paper.
\newpage
\centerline {\bf Appendix A }
\medskip
\medskip
The phase space factor [PS] in eq. (2) can be calculated
for different kinematic settings. In the c.m. system,
for the energy distribution of the outgoing protons it is,
\begin{equation}
[PS]=\frac {1}{(2\pi )^5}\frac {m_\Delta m^2E_AE_B}{\sqrt s}\frac
 {k_{p'}k_\pi ^3}{p_c}
\frac {1}{k_\pi ^2(\sqrt s-E_{p'})+E_\pi({\bf k}_{p'}.{\bf k}_\pi)}
dE_{p'}d\Omega _{p'}d\Omega _\pi
\end{equation}
For the mass distribution of the delta it is,
\begin{equation}
[PS]=\frac {\mu}{(2\pi)^5}\frac {m_\Delta m^2E_A E_B}{s}\frac
 {K_B k_\pi ^3}{p_c}
\frac {1}{k_\pi ^2(\sqrt s-E_B)+E_\pi ({\bf K}_B.{\bf k}_\pi)}
d\mu d\Omega _B d\Omega _ \pi,
\end{equation}
and for the four momentum transfer distribution it is given by
\begin{equation}
[PS]=\frac {\mu}{2(2\pi)^5} \frac {m_\Delta m^2E_AE_B}{s}\frac
{k_\pi ^3}{p_c^2}
\frac {1}{k_\pi ^2(\sqrt s-E_B)+E_\pi({\bf K}_B.{\bf k}_\pi)}
d\mu dt d\phi _B d\Omega _\pi .
\end{equation}

\newpage
\section *{REFERENCES}
\begin{enumerate}
\item see articles  in   Delta Excitation in Nuclei, ed. H.Toki et al.
(World Scientific, 1994);
A.\ B.\ Migdal, E.\ E.\ Sapirstein, M.\ A.\ Troitsky, and
D.\ N.\ Voskresensky, Phys. Rep. {\bf 192}, 179 (1990).
G. E. Brown, Phys. Rep. {\bf163}, 167 (1988);
B. K. Jain and A. B. Santra, Phys. Rep. {\bf 230}, 1 (1993);
 A. B. Migdal, Rev. Mod.
Phys. {\bf 50}, 10 (1978); C. Gaarde, Ann. Rev. Nucl. Science {\bf 41},
 187 (1991);
T.\ Udagawa, S.\ -W.\  Hong and F.\ Osterfeld,
Phys. Lett. B {\bf 245 }, 1 (1990); P.\ Oltmanns, F.\ Osterfeld, and
T.\ Udagawa, Phys. Lett. B {\bf 299}, 194 (1993);
T.\ Udagawa, P.\ Oltmanns, F.\ Osterfeld, and S.W.Hong
, Phys. Rev. C {\bf 49}, 3162 (1994);
C. Guet, M. Soyeur, J. Bowlin and G. E. Brown, Nucl. Phys. {\bf A494},
558 (1989);
P. Fernandez de Cordoba, E. Oset and M. J. Vicente-Vacas, 1994,
preprint;
H. J. Morsch et al., Phys. Rev. Lett. {\bf 69}, 1336 (1992);
T. Hennino
et al., Phys. Lett. B {\bf 283}, 42 (1992); ibid B {\bf 303},
236 (1993);
J. Chiba et al., Phys. Rev. Lett. {\bf 67}, 1982 (1991).
\item B. E. Bonner et al., Phys. Rev. C {\bf 18}, 1418 (1978);
C. Ellegaard et al., Phys. Rev. Lett. {\bf 50}, 1745 (1983):
C. Ellegaard et al., Phys. Lett. B {\bf 154}, 110 (1985);
D. Contardo et al., Phys. Lett. B {\bf 168}, 331 (1986);
I. Bergqvist et al., Nucl. Phys. {\bf A469}, 648 (1987).
\item T. Hennino et al., Phys. Rev. Lett. {\bf48}, 997 (1982).
\item V. G. Ableev et al., Pis'ma Zh. Eksp. Teor. Fiz. {\bf 40},
35 (1984) [JETP Lett. {\bf 40}, 763 (1984).
\item B. K. Jain, Phys. Rev. Lett. {\bf 50}, 815 (1983);
Phys. Rev. C {\bf 29}, 1396 (1984).
\item B. K. Jain, J. T. Londergan and G. E. Walker Phys. Rev.
 C {\bf 37}, 1564 (1988).
\item The $^{12}$C(p,p$' \pi ^{\pm}$) Reaction,  S. Yen et al.,
TRIUMF-Research Proposal $\#$ 636, 1994.
\item K. Gottfried and D. I. Julius, Phys. Rev. D {\bf  1}, 140 (1970).
\item G. Chapline, Phys. Rev. D {\bf 1}, 949 (1970); P. Osland and D.
Treleani, Nuclear Physics {\bf B107}, 493 (1976).
\item B. K. Jain and A. B. Santra, Phys. Lett. B {\bf 244}, 5 (1990);
Nucl. Phys. {\bf A519}, 697 (1990);
\item V.\ F.\ Dmitriev, O.\  Sushkov and C.\  Gaarde, Nucl. Phys.
{\bf A459}, 503 (1986).
\item Q. Haider and L. C. Liu, Phys. Lett. B {\bf 335}, 253 (1994).
\item D. V. Bugg, A. A. Carter and J. R. Carter, Phys. Lett.
 B {\bf 44}, 278 (1973); O. Dumbrajs et al., Nucl. Phys.
 {\bf B216}, 277 (1983);
V. Flamino, W. G. Moorhead, D. R. O. Morrison and N. Rivoire,
CERN Report CERN-HERA {\bf 83-01}, 1983.
\item Nuclear Sizes and Structure by D. F. Jackson and R. C. Barrett
{\em Clarendon Press, Oxford, 1977}; C. W. De Jager, H. De Vries, and
 C. De Vries, Atomic Data and Nuclear Data Tables {\bf 14}, 479 (1974).
\item N. G. Kelkar and B. K. Jain, Phys. Rev. C {\bf 46}, 845 (1992).
\item T. E. O Ericson and J. H$\ddot {u}$fner, Phys. Lett. B {\bf 33},
601 (1970).
\item P. C. Tandy, E. F. Redish and D. Bolle,
Phys. Rev. Lett. {\bf 35}, 921 (1975); Phys. Rev. C {\bf 16}, 1924 (1977).
\item M. Hirata, F. Lenz and K. Yazaki, Ann. Phys. (N.Y) {\bf 108},
116 (1977).
\item D. V. Bugg et al., Phys. Rev. {\bf 146}, 900 (1966);
S. Barshay, C. B. Dover and J. P. Vary, Phys. Rev. C {\bf 11},
360 (1975).
\item I. Abdul Jalil and D. F. Jackson, J. Phys. {\bf G}:Nucl. Phys.
{\bf 5}, 1699 (1979); R. M. Haybron
and H. McManus, Phys. Rev. B {\bf 136}, 1730 (1964);
P. G. Roos and N. S. Wall, Phys. Rev. {\bf 140}, B1237 (1965);
K. Seth,  Nucl. Phys. {\bf A138}, 61 (1969).
\item F. Shimizu et al., Nucl. Phys. {\bf A386}, 571 (1982):
{\bf A389}, 445 (1982).

\end{enumerate}

\newpage
\section*{Table I. Optical Potentials for Pion and Delta}
\begin{table}[htb]
\centering
\begin{tabular}{|c|c|c|c|}
\cline{1-4}
$T_\Delta  $ (MeV)& -$W_\Delta $ (MeV)&$T_\pi $ (MeV)&-$W_\pi $(MeV)\\
$<100$ & 45 &40 &25  \\
120 & 32 &100 & 67\\
200 & 35 &150 &88 \\
350 & 45 & 200 & 90 \\
450 & 48 & 250 & 57 \\
650 & 51 & 300 & 38 \\
800 & 52 & 350 & 33 \\
- &- & 400 & 28 \\
- &- & 600 & 26 \\
\cline{1-4}
\end{tabular}
\end{table}
\newpage
\section*{Figure Captions}
\begin{enumerate}
\item Projectile excitation diagram for the (p,p$'\pi ^+$) reaction.

\item Four momentum transfer distribution at 500 MeV beam energy for
$^{12}$C. Curve `a', no distortion for any continuum wave (PW results).
Curve
`b', includes distortions for the beam and delta. Curve `c', includes
distortions for beam, delta, p$'$ and $\pi ^+$.

\item Four momentum transfer distribution at 1 GeV beam energy for
$^{12}$C. The description of various curves is the same as in fig. 2.

\item Invariant mass distribution at 500 MeV beam energy for $^{12}$C.
$\theta _\Delta =0^0$. The description of various curves is the same as in
fig. 2.

\item Invariant mass distribution at 1 GeV beam energy for $^{12}$C.
$\theta _\Delta =0^0$. The description of various curves is the same as in
fig. 2.

\item The outgoing proton energy distribution in the (p,p$'\pi ^+$)
reaction for a co-planar geometry, $\theta _{p'}=10^0$ and
$\theta _\pi =-10^0$. The beam energy is
500 MeV and the target is $^{12}$C. The description of various curves
is the same as in fig. 2.

\item The outgoing proton energy distribution in the (p,p$'\pi ^+$)
reaction for a co-planar geometry, $\theta _{p'}=10^0$ and
$\theta _\pi =-10^0$. The beam energy is
1 GeV and the target is $^{12}$C. The description of various
curves is the same as in fig. 2.

\item Invariant mass distribution at 1 GeV beam energy for $^{208}$Pb.
$\theta _\Delta =0^0$. The description of various
curves is the same as in fig. 2.

\item The outgoing proton energy distribution in the (p,p$'\pi ^+$)
reaction for a co-planar geometry, $\theta _{p'}=10^0$ and
$\theta _\pi =-10^0$. The beam energy is
1 GeV and the target is $^{208}$Pb. The description of various
curves is the same as in fig. 2, except that the cross-sections in
curves `a' and `b' are shown after dividing them by factors of
100 and 10, respectively.

\item Sensitivity of the invariant mass distribution at 500 MeV
beam energy to the distortion of the delta in the nuclear medium.
The target nucleus is $^{12}$C and $\theta _\Delta =0^0$.
Curve `a', no delta distortion. Curve `b', delta distortion with
only imaginary part, W$_\Delta$, of its potential.
Curve `c', delta distortion with imaginary as well as real part of its
optical potential. All curves include the distortion (including the
real part of their potentials) of beam, p$'$ and $\pi ^+$.

\item Sensitivity of the invariant mass distribution at 1 GeV
beam energy to the distortion of the delta in the nuclear medium.
The target nucleus is $^{12}$C and $\theta _\Delta =0^0$. Description of
different curves is the same as in fig. 10.

\end{enumerate}

\end{document}